\documentclass[aps,pre,groupedaddress,showpacs]{revtex4}
\usepackage{amssymb}
\usepackage{graphicx}
\usepackage{epsf}
\usepackage{amsmath}
\usepackage{bm} 

\begin{document}

\title{Spatial Patterns Induced Purely by Dichotomous Disorder}
\author{J. Buceta and Katja Lindenberg}
\affiliation{
Department of Chemistry and Biochemistry, and Institute for Nonlinear
Science,
University of California San Diego, 9500 Gilman Drive, 
La Jolla, CA 92093-0340, USA}
\date{\today}

\begin{abstract}
We study conditions under which spatially extended systems with
coupling \emph{a la} Swift-Hohenberg exhibit spatial patterns induced
purely by the presence of quenched dichotomous disorder. Complementing
the theoretical results based on a generalized mean-field approximation,
we also present numerical simulations of particular dynamical systems that
exhibit the proposed phenomenology.
\end{abstract}

\pacs{05.40.-a, 47.54.+r, 05.10.Gg, 89.75.kD}

\maketitle

\section{Introduction}
\label{introduction}
Quenched disorder and dynamical disorder play an important
role in the properties of many physical systems.  Some examples that
illustrate this role in a variety of very different contexts
and support the wide interest in the subject of disorder
include the propagation of fronts in porous media~\cite{meakin},
the conduction properties
of doped semiconductors materials~\cite{solymar}, the shift of the Curie
temperature in ferromagnets~\cite{ma}, and the
so-called Anderson localization transition~\cite{anderson}.

At the same time, another broad field of perennial interest
is pattern formation~\cite{cross} because spatio-temporal structures
are omnipresent in the physical world~\cite{forexample}.
Our own recent contributions in this field involve the discovery of
novel
mechanisms for the appearance of spatio-temporal structures
upon periodic or random global alternation of pattern-free
dynamics~\cite{buceta1}.

Herein we coadunate
these two topics by investigating conditions under wich
the presence of quenched disorder is the mechanism that triggers
pattern formation.  Thus our interest lies in a particular type of
purely-noise-induced phenomenon: the appearance of patterns in disordered
systems which in an ordered state exhibit {\em no} pattern formation.
We concentrate on systems with quenched dichotomous disorder and
coupling term {\em a la} Swift-Hohenberg~\cite{swift}, although
one can easily envision generalizations of the
formalism to other kinds of disorder and/or coupling terms.

Noise-induced phenomena in spatially extended systems have been
particularly active areas of investigation in
the recent past~\cite{ojalvo}. Among these are phenomena
involving pattern formation
\emph{induced purely by fluctuations}~\cite{parrondo,zaikin,buceta2}.
The word ``purely'' here emphasises the fact that
the control parameter that determines the presence or absence of
patterns is the noise intensity.  Moreover,
if the noise term is replaced by a non-fluctuating parameter,
no spatio-temporal structures develop for \emph{any} value of that
parameter.  Our study differs from these previous ones
in that we consider quenched spatial disorder, that is, the fluctuations
have no temporal dependence.

Instead of focusing on a specific system at the outset, we explore
some general conditions under which these systems 
exhibit purely disorder induced patterns.
We then illustrate our findings with a family of systems that includes
the paradigmatic models of
noise-induced phase transitions~\cite{chris,mangioni,kim}
and noise-induced spatial patterns~\cite{chris,parrondo}.
We also identify a phenomenon in our system that has previously only
been identified  in systems with colored
noise~\cite{mangioni,kim,buceta3}
or field-dependent kinetic coefficients~\cite{buceta2}, namely, a
re-entrant behavior with increasing coupling strength. In other words,
we find that increasing the coupling strength leads to
non-monotonic behavior such that the patterns are most
prominent for a finite value of the coupling and disappear altogether
when coupling is too strong (or too weak).  As in
second and first order phase transitions in equilibrium systems,
the re-entrance phenomenon can be either continuous (second order
behavior) or it may present multi-stability and associated
hysteresis (first order behavior).

The paper is organized as follows. We introduce the formalism in
Sec.~\ref{secmodel}. In Sec.~\ref{modulated} and Appendix~\ref{a} we
present a
generalized mean field approximation and state the requirements
for pattern formation induced purely by disorder.
The possible behaviors that can be deduced from these requirements
are explored in Secs.~\ref{behaviors} and~\ref{features}.
Particular examples of systems that exhibit pattern formation
are given in Sec.~\ref{examples}, and the order parameters used to
characterize the patterns are introduced and related to one another
in Appendix~\ref{b}.
Numerical simulations that confirm the qualitative validity of the
theoretical results are presented in Sec.~\ref{numerical}. Finally, we
summarize the main results in Sec.~\ref{conclusions}.

\section{The Model}
\label{secmodel}

We consider the following stochastic dynamics for a scalar field 
$\phi_{\bm r}\equiv\phi({\bm r},t) $ in the presence of
dichotomous disorder: 
\begin{equation}
\dot{\phi}_{\bm r}=f(\phi_{\bm r}) +g(\phi_{\bm r})
\xi_{\bm r}+\mathcal{L}\phi_{\bm r}.
\label{model}
\end{equation}
Here $\xi_{\bm r}$
is a space dependent \emph{quenched} dichotomous variable that
models spatial disorder. The probability
density of $\xi $ reads: 
\begin{equation}
\rho (\xi) =p_+\delta(\xi -\Delta)
+p_-\delta (\xi +\Delta),
\end{equation}
that is, at any given site ${\bm r}$ the variable $\xi_{\bm r}$
takes on either the value $+\Delta $ or the value $-\Delta $ with
probabilities $p_+$ and $p_-$ respectively. The term
$\mathcal{L}$ stands for
the
Swift-Hohenberg coupling operator
\begin{equation}
\mathcal{L=}-D(k_0^2+\nabla^2)^2.
\end{equation}
The effect of this coupling can be deduced by applying $\mathcal{L}$ to a
plane wave $e^{i{\bm k}\cdot{\bm r}}$,
\begin{equation}
{\cal L}e^{i{\bm k}\cdot{\bm r}}=\omega ( k)
 e^{i{\bm k}\cdot{\bm r}},
\end{equation}
where $\omega (k) =-D(k_0^2-k^2)^2$ is the continuous dispersion relation
(we use bold for vectorial quantities and italic for their magnitudes).

In order to implement a mean field theory for this system we need to
distinguish from one another neighboring locations ${\bm r}$ and
${\bm r}'$, which in turn
requires that we discretize the system.  Since numerical simulations also
involve discretization, this procedure does not interfere with the
comparisons of theoretical and numerical results.  With the understanding
of the action of the translation operator 
\begin{equation}
\exp \left(\delta x\frac{\partial }{\partial x}\right)f(x) =f(x+\delta x),
\end{equation}
it is straightforward to deduce a discrete version of the
Swift-Hohenberg coupling operator $\mathcal{L}$,
\begin{equation}
\mathcal{L}=-D\left[ k_{0}^{2}+\left(\frac{2}{\Delta x}\right)^2
\sum_{i=1}^{d}
\sinh^2\left( \frac{\Delta x}{2}\frac{\partial }{\partial x_{i}}\right)
\right] ^{2},
\label{l-discrete}
\end{equation}
where $d$ stands for the spatial dimension, $\Delta x$ for the lattice
spacing, and $\frac{\partial }{\partial x_i}$ indicates a partial
derivative with respect to component $i$ of the position vector
$\bm r =(x_1,x_2,\ldots ,x_i,\ldots ,x_d ) $.
The continuum delta function
$\delta({\bm r}-{\bm r}')$ is replaced in the usual way by a ratio
that contains the Kronecker delta and the lattice spacing,
$\delta_{{\bm r},{\bm r}'}/(\Delta x)^d$.  As in the
continuous case, the discrete
dispersion relation can be obtained by applying
the operator (\ref{l-discrete}) to a plane wave
$e^{i{\bm k}\cdot{\bm r}}$, to obtain
\begin{equation}
\omega({\bm k}) =-D\left[k_0^2-\left(\frac{2}{\Delta x}\right)^2
\sum_{i=1}^d \sin^2 \left(\frac{\Delta x}{2}k_i \right)\right]^2.
\label{dispersion}
\end{equation}
Here $k_i$ denotes component $i$ of the wave vector ${\bm k}
=(k_1,k_2,\ldots,k_i,\ldots,k_d)$.

Note that $\omega({\bm k})$ is nonpositive for any value of $k$ in both the
continuous and discrete cases, and that in the discrete case it depends
not only on the magnitude but also on the direction of ${\bm k}$. 
Of particular importance in our subsequent analysis are those modes  
for which $\omega({\bm k})=0$.  In the continuum these are the modes with
$k=k_0$, which are all those that lie on a continuous hypersurface of radius 
$k_0$ around the origin.  In the discretized system the magnitudes
$k^*$ of the least stable modes are shifted from $k_0$ and depend on
direction, as can be seen by solving Eq.~(\ref{dispersion}).  The longest
vectors such that $\omega({\bm k}^*)=0$ lie along the cartesian directions in
${\bm k}$ space, e.g. $(k^*,0,0,\ldots, 0)$ and have magnitude
\begin{equation}
\max k^{\ast}=\frac{2}{\Delta x}\arcsin\left(\frac{k_0\Delta x}{2}
\right).
\label{mink}
\end{equation}
The shortest lie along a reciprocal space diagonal, e.g.
$\frac{1}{\sqrt{d}}(k^*,k^*,k^*,\ldots,k^*)$, and have magnitude
\begin{equation}
\min k^{\ast}=\frac{2\sqrt{d}}{\Delta x}\arcsin\left(
\frac{k_0\Delta x}{2\sqrt{d}}\right).
\label{maxk}
\end{equation}
In our analysis in $d=2$ we take $k_0=1$ and $\Delta x=1$. 
The difference between
$\max k^{\ast}$ and $\min k^{\ast}$ is smaller than 3\%,
the two values being $\pi/3 = 1.0472$ and $\pi/3.0737 = 1.0221$.
Therefore it is only a mild approximation to neglect the directional
dependence of the solutions of $\omega({\bm k}^*)=0$ and focus on the
magnitude, $\omega(k^*)=0$.  Furthermore, in
simulations one must use a {\em finite} system of $N^d$ sites [i.e., of
volume $(N\Delta x)^d$], so that the allowed
modes themselves form a discrete set, with each component separated from the
next one by an interval $\delta k = 2\pi/N \Delta x$. 
One way to pick the least stable modes is to construct a ring
of radius $\langle k^*\rangle$ (which we shall simply call
$k^*$ from here on) of
thickness $\delta k$ and to consider all the modes that lie in this ring.
We can then estimate the number of modes $\mathfrak{n}(k^*)$ that are
least stable by calculating the number of cells of volume
$(2\pi/N)^d$ in the ring:
\begin{equation}
\mathfrak{n}(k^{\ast}) =\frac{d\pi^{d/2}}{\Gamma(d/2+1)}
\left(\frac{N\Delta x  k^*}{2\pi}\right)^{d-1}.
\label{frakn}
\end{equation}
Although slight variations in the particular way of counting are
possible, for sufficiently large $N$ the differences are small.

\section{Modulated Mean-Field Theory}
\label{modulated}

To establish the existence of patterns of a characteristic length
scale in the steady state, we seek a spatially periodic structure defined by
wavevectors whose magnitude is associated with the inverse of this
length scale.  The appropriate wavevectors to focus on
are precisely those of magnitude $k^*$, that is, those for which
$\omega({\bm k})=0$.  A mean field theory requires that we make
an ansatz about the behavior of the stationary field at sites
${\bm r}' \neq {\bm r}$ which are coupled to the focus site ${\bm r}$
by the operator ${\mathcal L}$. We require that the ansatz capture
correct limiting behaviors and also
incorporate an appropriate spatial modulation.  Our choice is
\begin{equation}
\phi_{\bm r^\prime}={\mathcal A}(k^*) \sum_{\left\{\bm{k}^{\ast}\right\}
}
\cos \left[
{\bm k}\cdot ({\bm r}-{\bm r}^\prime) \right] + B,
\label{mean field}
\end{equation}
where the sum (or, in an infinite system, the integral) is
over wavevectors of magnitude $k^*$.  The constant $B$ is specified
below.  Our ansatz incorporates the assumption that all modes of 
magnitude $k^\ast$ contribute
with equal (direction-independent) weight ${\mathcal A}(k^*)$.
In Appendix~\ref{a} we show that the action of the
coupling operator on this ansatz is given by
\begin{equation}
\mathcal{L}\phi_{\bm{r}}=D_1\left[\mathfrak{n}(k^{\ast})
\mathcal{A}(k^{\ast})-\phi_{\bm{r}}\right] +B(D_1-Dk_0^4).
\label{finalapprox}
\end{equation}
where
\begin{equation}
D_1=D\left[\left(\frac{2d}{(\Delta x)^2}
-k_0^2\right)^2+\frac{2d}{(\Delta x)^4}\right].
\label{d1}
\end{equation}
Substitution of Eq.~(\ref{finalapprox})
in Eq.~(\ref{model}) then leads to an equation that depends
only on a generic site index ${\bm r}$ that can simply be dropped:
\begin{equation}
\dot{\phi}=f(\phi) +g(\phi)\xi +D_1\left[ \mathfrak{n}({k^\ast})
\mathcal{A}(k^\ast) -\phi \right] +B(D_1-Dk_0^4).
\label{ec1}
\end{equation}

In the steady state we can write the explicit equations associated with
Eq.~(\ref{ec1}) as
\begin{eqnarray}
0&=&F_+(\phi)
+D_1 \left[ \mathfrak{n}({k^\ast})
\mathcal{A}(k^\ast) -\phi \right] +B(D_1-Dk_0^4), \nonumber \\ \nonumber \\
0&=&F_-(\phi)
+D_1 \left[ \mathfrak{n}({k^\ast})
\mathcal{A}(k^\ast) -\phi \right] +B(D_1-Dk_0^4), 
\label{ec1r}
\end{eqnarray}
where we have introduced the shorthand notation
\begin{equation}
F_{\pm }(\phi) =f(\phi) \pm g(\phi)\Delta.
\label{short}
\end{equation}
We denote the solutions of Eqs.~(\ref{ec1r}) 
by $\phi_\pm$ respectively.

The amplitude ${\mathcal A}(k^*)$ and the constant $B$ are the
mean field quantities that must be chosen self-consistently to
complete the solution of the problem. To close these equations we choose
\begin{equation}
B=p_+\widetilde{\phi}_+ + p_-\widetilde{\phi}_-
\label{B}
\end{equation}
where the $\widetilde{\phi}_\pm$ are the steady state fields for each
of the separate dynamics in completely ordered systems, that is,
one with $p_+=1$ and one with $p_-=1$ (see below).
Furthermore, we impose the self-consistency condition
\begin{equation}
[\mathfrak{n}({k^\ast})\mathcal{A}(k^\ast) +B] = \langle \phi \rangle
=\int_{-\infty}^\infty  d\phi \;
\phi \rho(\phi ;\mathcal{A}(k^\ast)) d\phi . 
\label{selfcon}
\end{equation}
Since 
\begin{equation}
\rho (\phi ;\mathcal{A}(k^\ast)) =
p_+\delta(\phi -\phi _{+}) +p_- \delta(\phi-\phi_{-}),
\end{equation}
the self-consistency condition can be rewritten as
\begin{equation}
\mathfrak{n}({k^\ast})\mathcal{A}(k^\ast) =p_+(\phi _{+}-\widetilde{\phi}_+)
+p_-(\phi _{-}-\widetilde{\phi}_-),
\label{selfcon2}
\end{equation}
where it should be stressed that the $\phi_\pm$ are of course functions of
$\mathcal{A}(k^\ast)$.
Finally, Eqs.~(\ref{ec1r}) can then be rewritten as
\begin{equation}
0=F_{\pm }(\phi _{\pm }) +D_1p_{\mp}
(\phi_{\mp }-\phi_{\pm }) -Dk_0^4 (p_+\widetilde{\phi}_+ +
p_-\widetilde{\phi}_-).
\label{ec3b}
\end{equation}

We are particularly interested in systems in which there are \emph{no
patterns} in any ordered state, that is, where patterns are purely
disorder-induced (we will choose the functions $f$ and $g$ accordingly).
For each value of the dichotomous disorder parameter we can write
a deterministic evolution equation for the dynamics.
We insist that each of these evolution equations describe a system that
intrinsically has at least \emph{one} steady state and hence we insist
that the associated forces be confining.
We also require that each of these dynamics be pattern-free, so that
there must be exactly \emph{one} steady state solution for each, and this
steady state solution in each case must be a constant independent of
$\bm{r}$.  The confining condition requires that for all $\Delta$
\begin{eqnarray}
\underset{\phi \rightarrow -\infty }{\lim }F_\pm(\phi)-Dk_0^4\phi
&=& \infty \nonumber\\
\underset{\phi \rightarrow \infty }{\lim }F_\pm(\phi)-Dk_0^4\phi
&=& -\infty .
\end{eqnarray}
Indeed, it is reasonable to require that the confinement not be due
simply to the coupling.  and thus to require that 
\begin{equation}
\underset{\phi \rightarrow -\infty }{\lim }F_\pm(\phi)
= \infty, \qquad 
\underset{\phi \rightarrow \infty }{\lim }F_\pm(\phi)
= -\infty .
\label{conditions}
\end{equation}
The steady state conditions follow directly from Eq.~(\ref{model}):
\begin{equation}
0=F_{\pm }(\phi) - Dk_0^4\phi . 
\label{ec3a}
\end{equation}
The solutions [already introduced in the choice of $B$ in Eq.~(\ref{B})]
are denoted by $\widetilde{\phi}_+$ and
$\widetilde{\phi}_-$ respectively. Further, to avoid any insertion of
patterns other than those induced by disorder, we also insist
that $f(\phi)$ itself be associated with only a single steady state,
i.e., that the equations $f(\phi)-Dk_0^4\phi=0$ and $f(\phi)=0$ 
also have only a single
solution.  On the other hand, if we approximate
the dynamics in these ordered systems by our
mean field ansatz, and insist that $\mathcal{A}(k^\ast)=0$ since there
are no patterns, the equations in the steady state with $B$ chosen
as in Eq.~(\ref{B}) in each case read
\begin{equation}
0=F_{\pm }(\phi) -D_1 \phi +\widetilde{\phi}_\pm(D_1-Dk_0^4) . 
\label{ec3}
\end{equation}
Clearly, the solutions of 
the two equations are again $\widetilde{\phi}_+$ and $\widetilde{\phi}_-$
respectively, and thus our choice of $B$
is consistent with the exact unpatterned solutions in the ordered steady
states.  

\begin{figure}
\begin{center}
\includegraphics[width = 8cm]{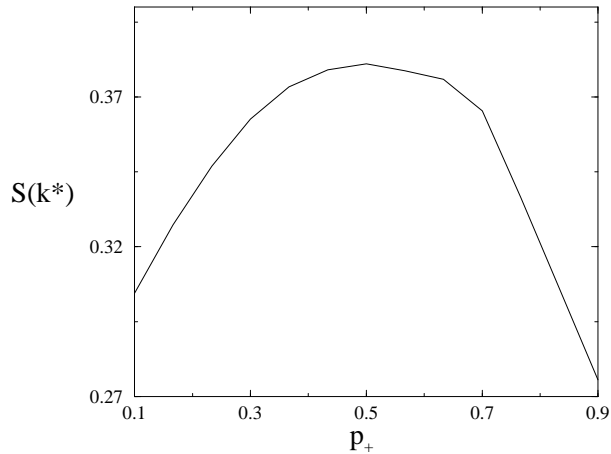}
\end{center}
\caption{Order parameter vs $p_+$ for a model detailed in
Sec.~\ref{examples}.  Pattern formation is associated with a nonzero
value of the order parameter.  
}
\label{p}
\end{figure}

For simplicity, in most of the remainder of this work we set
$p_+=p_-=1/2$, although we hasten to add that this condition is not
necessary for the appearance of patterns.  To support this statement, we
show in Fig.~\ref{p} the order parameter
$S(k^\ast)=\mathfrak{n}(k^\ast)\mathcal{A}^2(k^\ast)$ (discussed in more
detail in Sec.~\ref{examples} and Appendix~\ref{b}) vs $p_+$ for a
particular model considered later; the details of that model are not
important at this point.  A nonzero value of the order
parameter indicates the appearance of patterns, and a higher value is
indicative of more pronounced patterns.  In this particular instance the
strongest patterns occur when $p_+=1/2$, but the figure shows that
other values of $p_+$ also
lead to pattern formation for the same model and parameter values.
Furthermore,
we also impose the
requirement that $f(\phi)$ be an odd function and $g(\phi)$ an even
function of $\phi$. This immediately leads to the symmetry
\begin{equation}
F_+(\phi_+)=-F_-(-\phi_-)
\end{equation}
and, associated with the pattern-free solution $\mathcal{A}(k^*)=0$, 
\begin{equation}
\widetilde{\phi}_+=-\widetilde{\phi}_-
\end{equation}
In particular, the average homogeneous solution $B=0$. This is always
one of the possible solutions of the problem.  The stability of this
state, and the possible appearance of other solutions with
$\mathcal{A}(k^*)\neq 0$, are the subjects of our further analysis.

We collect, then, the system of equations to be analyzed with these
simplifications.  Furthermore, taking advantage of the symmetries
established so far, the equations take on a more tractable appearance if
we define new variables
\begin{equation}
x\equiv \frac{D_1}{2}\phi_+, \qquad y\equiv -\frac{D_1}{2}\phi_-, \qquad 
x_0=\frac{D_1}{2}\widetilde{\phi}_+ =-\frac{D_1}{2}\widetilde{\phi}_-.
\end{equation}
Thus the equations to solve are [cf. Eq.~(\ref{ec1r})]
\begin{eqnarray}
F(x)-x&=&y\nonumber\\ [5pt]
F(y)-y&=&x
\label{thesearetheones}
\end{eqnarray}
with
\begin{equation}
D_1\mathfrak{n}({k^\ast})\mathcal{A}(k^\ast) = x-y,
\label{theproof}
\end{equation}
where we have applied the symmetry $F_+(x)=-F_-(y)$ and have dropped the
subscript $+$ since we only need to use $F_+$:
\begin{equation}
F(x)\equiv F_+(x) = f(x) +g(x)\Delta.
\end{equation} 
To this we add the requirement that follows from Eq.~(\ref{conditions}),
\begin{equation}
\underset{x \rightarrow \pm\infty }{\lim }F^\prime(x)<0,
\label{asymptotic}
\end{equation}
where the prime denotes a derivative with respect to the argument.
The solution $x=y=x_0$ is pattern-free.  We
seek solutions with $x\neq y$ to establish the appearance of
patterns.  Note that if a pair $(x,y)$ solves
Eqs.~(\ref{thesearetheones}), then so does the pair $(y,x)$, simply
leading to a reversal in the sign of $\mathcal{A}(k^\ast)$.  Since
negative values of $\mathcal{A}(k^\ast)$ can be interpreted in terms of
an overall spatial phase, the second pair adds no new physical
information beyond the symmetry statement.

The stability of the pattern-free solution becomes an important issue in
our further discussion. This can be established by a linear stability
analysis around $x=y=x_0$. The time evolution of small
perturbations about this solution is obtained by expanding the evolution
equations
\begin{equation}
F(x)-x-y =\dot{x},\qquad
F(y)-y-x = \dot{y},
\end{equation}
at $(x,y)=(x_0+\delta x, x_0+\delta y)$ and retaining terms up to
first order in the perturbations:
\begin{equation}
\left(
\begin{array}{c}
\delta \dot{x}\\ [10pt]
\delta \dot{y}
\end{array} 
\right)
= 
\left(
\begin{array}{cc}
F^\prime(x_0)-1       &-1\\ [10pt]
-1& F^\prime(x_0)-1
\end{array}
\right) \left(
\begin{array}{c}
\delta x\\ [10pt]
\delta y
\end{array}
\right).
\end{equation}
The $2\times 2$ evolution matrix has eigenvalues
\begin{equation}
\lambda_+ = F^\prime(x_0), \qquad \lambda_-=F^\prime(x_0) -2.
\end{equation}
The largest eingenvalue (largest Lyapunov exponent) is clearly
$\lambda_+$, and it determines the stability of the pattern-free
solution:
\begin{equation}
\begin{array}{ll}
F^\prime(x_0) < 0 &\mbox{    solution is stable}\\ [10pt]
F^\prime(x_0) > 0 &\mbox{    solution is unstable}.
\end{array}
\label{diagnosis}
\end{equation}

\section{Pattern Formation Induced Purely by Disorder}
\label{behaviors}

\begin{figure}[htb]
\begin{center}
\epsfxsize = 2.in
\epsffile{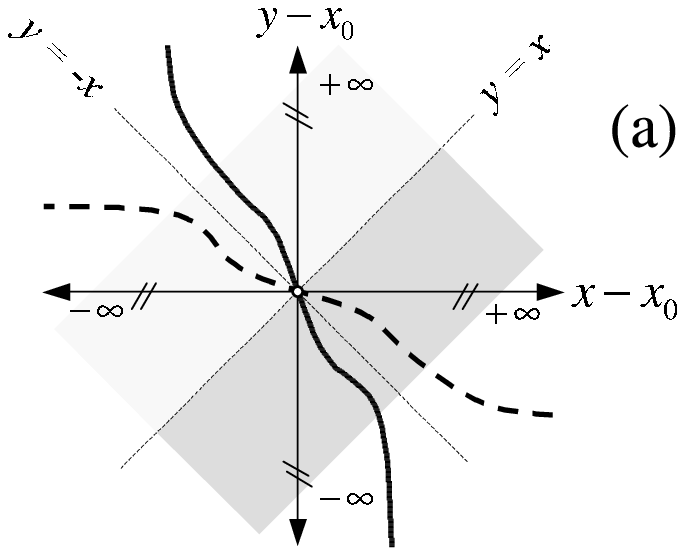}
\epsfxsize = 2.in
\epsffile{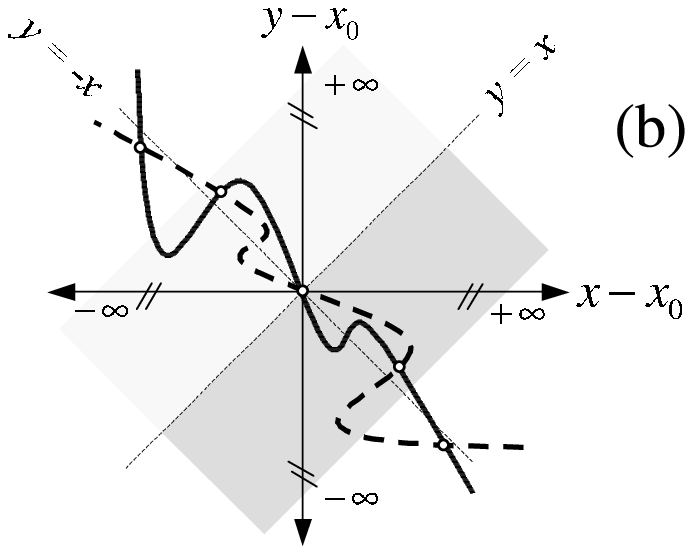}
\epsfxsize = 2.in
\epsffile{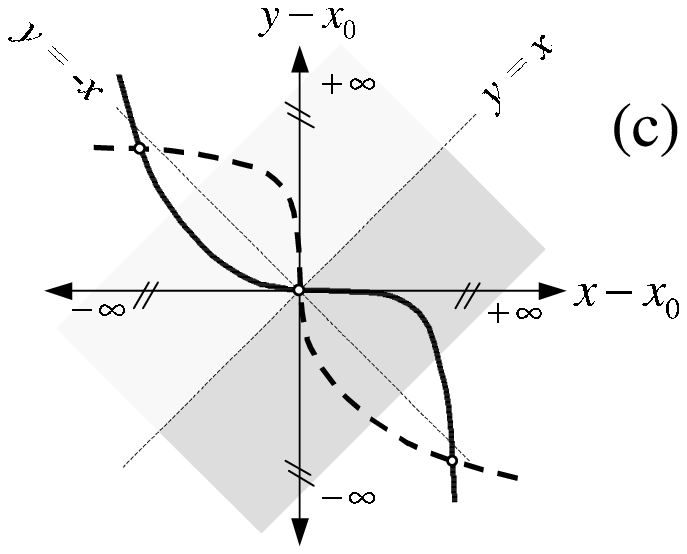}
\end{center}
\caption{
Schematic of possible solutions of the mean field equations.  The origin
represents the pattern-free solution.  Other intersections of the
thick solid curve (which represents $F(x)-x$) and
the thick dashed curve (which represents $F(y)-y$) are solutions of 
Eq.~(\ref{thesearetheones}) that lead to patterns.  Panel (a) represents
a case in which the pattern-free solution is stable and there
are no other solutions, that is, there is no pattern formation.  Panel (b) 
describes the coexistence of the stable pattern-free state with
a patterned stable state.  In panel (c) the pattern-free state is unstable
and the only stable state is a patterned state.
}
\label{f1}
\end{figure}

We seek solutions for Eqs.~(\ref{thesearetheones}) and, in
particular, solutions with $x\neq y$. For given potential functions,
one
could produce three-dimensional plots of $F(x)-x-y$ and
$F(y)-y-x$ vs $x$ and $y$ and observe the intersections of these
surfaces.  A more intuitive graphical way to organize this search is
presented in the three panels of
Fig.~\ref{f1}.  The origin represents the pattern-free solution, and
this is the only point on the line $y=x$ that solves the equations
since $F(x)$ is not an odd function.
Since Eqs. (\ref{thesearetheones}) are invariant under the
transformation $x\longleftrightarrow y$,
the line $x=y$ defines a specular plane, that is, $F(y)-y$
is a specular image of $F(x)-x$ with respect to that
symmetry plane.  
The asymptotic behavior (\ref{asymptotic}) tells us that for
sufficiently large $|x|$ in the upper left quadrant, $F(x)-x$ lies
above the line $x=-y$ and has a slope $<-1$. This is schematically
indicated by the thick solid line in
the upper left quadrant of each panel in the figure.  Similarly,
in the lower right
quadrant the thick solid line recognizes that $F(x)-x$ must lie below
the line $x=-y$ with a slope $<-1$.  The specular symmetry around the
line $x=y$ then leads us to the asymptotic thick dashed lines representing
the behavior of $F(y)-y$. 

In panel (a) we
illustrate a case for which the pattern-free solution is
\emph{stable}, that is, $F^\prime(x_0)<0$.  This is indicated by
the thick solid line going through the origin.  Again, the specular
symmetry leads us to the thick dashed line to indicate the approriate
slope for $F(y) -y$.  Now it is clear that the thick solid lines can be
connected, and the thick dashed lines can be connected, in such a way
that the two lines do not cross anywhere else but at the origin.  Thus,
when the state $(x_0,x_0)$ is stable, there may not appear any other
stationary states and the system may simply be pattern free.  This is
the case we have sketched in panel (a). 

On the other hand, it is
possible to connect the solid and dashed curves respectively in such a
way that there are \emph{two} additional crossings of the curves
(actually four crossings, but only two provide independent information).
These represent two additional steady state solutions, each leading to
pattern formation since $x\neq y$.  However, only one of the two
is stable.  This is illustrated in
panel (b). This represents the case of \emph{coexisting} stable states,
one pattern-free and the other patterned, separated by an unstable state.
Such coexisting states are characteristic of
\emph{first order phase transitions}.  One can
carry this further and envision further crossings, always an \emph{even}
number of additional crossings, representing stable patterned states
that coexist with one another and with the pattern-free state, separated
by unstable states.  

In panel (c) we illustrate a case for which the pattern-free solution is
\emph{unstable}, that is, $F^\prime(x_0)<0$.  Again, this is indicated
by the thick solid line going through the origin together with its
dashed specular partner.  Now it is clear that a connection of the lines
necessarily leads to at least one crossing with $x\neq y$.  In other
words, when the pattern-free state becomes unstable, at least one
patterned state is necessarily stable.  \emph{Therefore, a
sufficient condition for the occurrence of patterns is that the
pattern-free state become unstable}.

In summary, we have found the following general behavior:
\begin{itemize}
\item
When the pattern-free solution is stable, there may or may not occur
one (or more) stable solution(s) that leads to pattern formation, the
patterned and unpatterned stable states being
separated from one another by unstable solutions.  The appearance of
such an \emph{additional} stable solution is indicative of 
a \emph{first order phase transition}, with the usual coexistence
and hysteresis characteristics.  
\item On the other hand, when the
pattern-free solution is unstable, a patterned stable state necessarily
appears.  The destabilization of the pattern-free state might
simply mark the end of the coexistence region of a first-order
transition described above, or
it might mark the occurrence of a second order transition.  These
alternatives are discussed in further detail below.
\end{itemize}

We note here that although we have not said so explicitly, it is
implicit in this entire analysis that patterned states can only exist if
the potential functions are {\em nonlinear}.  Linear forms can not
satisfy the conditions that lead to the emergence of patterns.

\section{Mean Field Solution - General Features}
\label{features}

\begin{figure}[htb]
\begin{center}
\epsfxsize = 2.in
\epsffile{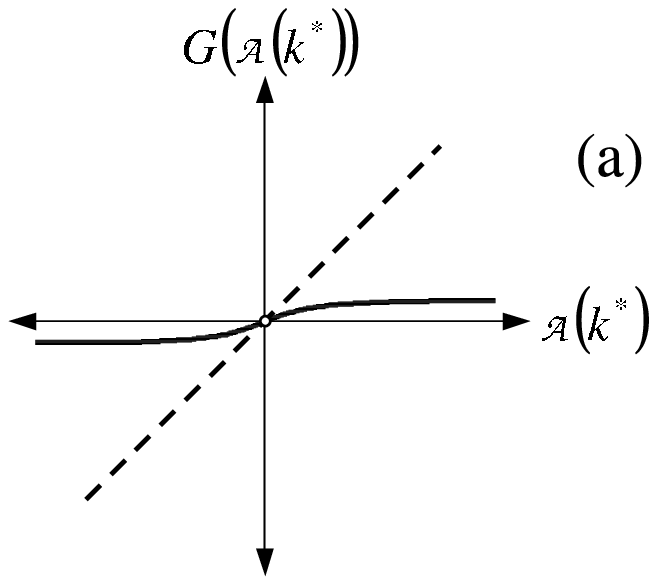}
\epsfxsize = 2.in
\epsffile{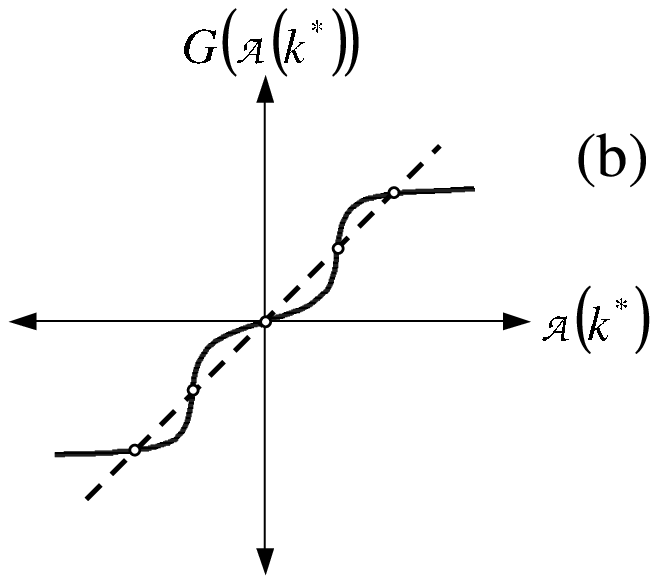}
\epsfxsize = 2.in
\epsffile{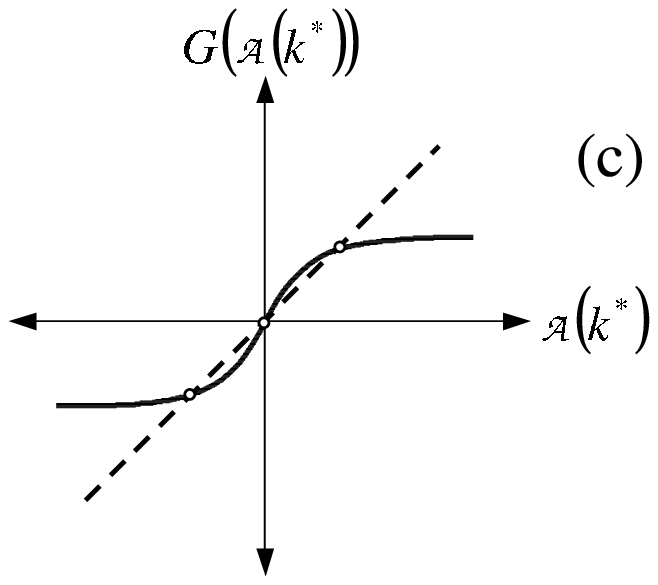}
\end{center}
\caption{
Schematic of $G(\mathcal{A})$ vs $\mathcal A$. Intersections of the
thick solid curve and the thick dashed line represent solutions 
of the mean field equations.  Each panel represents the same situation
as the corresponding panel in Fig.~\ref{f1}.  Panel (a) depicts the case
of a stable pattern-free solution and no other.  Panel (b) shows the
case of coexistence of the stable pattern-free state and a patterned
stable state (separated by an unstable solution).  In panel (c) the
pattern-free state is unstable and only a patterned state is stable.
}
\label{f2}
\end{figure}

Another way to exhibit the variety of possible transitions discussed
above is to focus on the explicit solution of the mean field equations 
and consider the resulting values of $\mathcal{A}(k^\ast)$. We start with
the first Eq.~(\ref{thesearetheones}) and subtract $x$ from both sides,
$F(x)-2x = y-x$.  Similarly, we subtract $y$ from both sides of the
second equation, to write $F(y)-2y = x-y$.  Using Eq.~(\ref{theproof})
then implies that we can write
\begin{eqnarray}
D_1\mathfrak{n}({k^\ast})\mathcal{A}(k^\ast)
&=& -F(x)+2x \equiv H_-(x)
\nonumber\\
&=& F(y)-2y \equiv H_+(y).
\end{eqnarray}
We can invert these relations, 
\begin{equation}
x=H_-^{-1}(\mathcal{A}(k^\ast)), \qquad
y=H_+^{-1}(\mathcal{A}(k^\ast)).
\end{equation}
The self consistency condition (\ref{theproof}) for the mean field
solution can then be written as
\begin{equation}
D_1\mathfrak{n}({k^\ast})\mathcal{A}(k^\ast)
=\left[H_-^{-1}(\mathcal{A}(k^\ast) )
     -H_+^{-1}(\mathcal{A}(k^\ast)) \right] 
\equiv D_1\mathfrak{n}({k^\ast})G(\mathcal{A}(k^\ast)) .
\label{self-con3}
\end{equation}
Since $\mathcal{A}(k^\ast)=0$ is always a solution (albeit not always
stable), we know that $G(0)=0$. Furthermore, the symmetry of the problem
implies that if $\mathcal{A}(k^\ast)$ is a solution,  then so is
$-\mathcal{A}(k^\ast)$.  Therefore
$G(\mathcal{A}(k^\ast))$ is an odd function.
An expansion about zero thus has only odd powers,
\begin{equation}
G(\mathcal{A}(k^\ast)) = a\mathcal{A}(k^\ast)
+b\mathcal{A}^3(k^\ast)
+\mathcal{O}\left(\mathcal{A}^5(k^\ast)\right).
\label{taylor}
\end{equation}
Our analysis proceeds on the basis of the first term of this expansion
as well as the asymptotic behavior of the function $G$. 
The graphical representation of this analysis is shown in the three
panels of Fig.~\ref{f2}, each associated with the corresponding panels
in Fig.~\ref{f1}.  

Let us first deduce the asymptotic behavior of $G$ by considering the
slope 
\begin{equation}
\frac{\partial H_-^{-1}(\mathcal{A})}{
\partial \mathcal{A}}
=\frac{1}{\frac{\displaystyle d}{\displaystyle
dx}H_-(x)}
=\frac{1}{\frac{\displaystyle d}{\displaystyle dx} [-F(x)+2x]} =
\frac{1}{[2-F^\prime(x)]},
\end{equation}
where we have applied the general relation 
\begin{equation}
\frac{d}{dz}h^{-1}(z) = \frac{1}{h^\prime [h^{-1}(z)]}
\end{equation}
and the prime, as usual, denotes a derivative with respect to the
argument.  In particular, we thus find that
\begin{equation}
\underset{\mathcal{A} \rightarrow \pm\infty }{\lim }
\frac{\partial H_-^{-1}(\mathcal{A})}{
\partial \mathcal{A}}
= \underset{x \rightarrow \pm\infty }{\lim }
\frac{1}{[2-F^\prime(x)]} <\frac{1}{2},
\end{equation}
where we have used Eq.~(\ref{asymptotic}).
The slope $-\partial H_+^{-1}(\mathcal{A})/\partial \mathcal{A}$ leads
to exactly the same asymptotic result, so that it follows from 
Eq.~(\ref{self-con3}) that
\begin{equation}
\underset{\mathcal{A} \rightarrow \pm\infty }{\lim }\frac{\partial 
G(\mathcal{A}(k^\ast))}{\partial \mathcal{A}(k^\ast)} <1.
\end{equation}
This clearly implies that asymptotically the function $G(\mathcal{A})$
in the positive half-plane must lie below the diagonal line, as we
have drawn in the three panels in Fig.~\ref{f2} (in the negative half
plane it must lie above, again as shown).
We keep in mind that patterned solutions are associated with 
intersections of the function $G(\mathcal{A})$ and the diagonal,
away from the origin.

Next we look at the behavior of $G$ near the origin, and consider 
the coefficient $a$ in the Taylor expansion (\ref{taylor}). 
The calculation of this coefficient involves
precisely the steps followed above, but with the functions evaluated at
$x_0$ instead of asymptotically.  We readily obtain
\begin{equation}
a\equiv  
\left. \frac{\partial G(\mathcal{A}(k^\ast)}{\partial \mathcal{A}(
k^\ast)}\right|_{\mathcal{A}(k^\ast) =0} = \frac{2}{2-F^\prime(x_0)}.
\end{equation}
Now suppose that $F^\prime(x_0)<0$, a condition that according to
Eq.~(\ref{diagnosis}) means that the pattern-free solution is stable.
The slope of $G(\mathcal{A}(k^\ast))$ near the origin is below the
diagonal, $a<1$.  It is then possible that
other than the crossing at the
origin there is no other crossing, that is, only the stable pattern-free
solution exists.  This is illustrated in panel (a) of Fig.~\ref{f2}, and
corresponds to the situation in panel (a) of Fig.~\ref{f1}. 

Another
possibility is that there are an even number of additional crossings, as
illustrated in panel (b).  One of the two additional solutions
shown in the panel would
be stable, the other unstable, and the stable patterned solution
would coexist with the pattern-free solution
under the circumstances shown in the panel.  Again, this corresponds to
the situation in panel (b) of Fig.~\ref{f1}.  Below we establish further
conditions on the forces that might lead to this behavior.  

Next suppose that $F^\prime(x_0)>0$, the condition that according to 
Eq.~(\ref{diagnosis}) is associated with an unstable pattern-free solution. 
The slope of $G(\mathcal{A}(k^\ast))$ near the origin is now above the
diagonal in the positive half-plane, $a>1$, and another crossing than the
one at the origin certainly occurs, thus insuring a stable patterned solution.
This is shown in panel (c) of Fig.~\ref{f2} and corresponds to 
panel (c) in Fig.~\ref{f1}.   

Although neither a necessary nor a sufficient condition,
it is apparent that the additional crossings in panel (b) of
Fig.~\ref{f2} might be accompanied by a positive 
curvature of $G(\mathcal{A})$ near the origin, as shown in the panel,
that is, additional crossings might occur if the
coefficient $b$ in Eq.~(\ref{taylor}) is positive.  Likewise, case (c) is
likely to be associated with a negative curvature near the origin,
$b<0$.

It is useful to take cognizance of the possible sequences of behavior
as one varies model parameters.  For instance, a
sequence (a) $\rightarrow$ (b) $\rightarrow$ (c) would signal a first
order phase transition from an unpatterned state through a coexistance
regime of unpatterned and patterned states to a regime where only the
patterned state is stable.  On the other hand, a sequence (a)
$\rightarrow$ (c) would represent a second order phase transition from
an unpatterned to a patterned state.  Transitions might be re-entrant,
so that a returns (c) $\rightarrow$ (b) $\rightarrow$ (a) (first order) or 
(c) $\rightarrow$ (a) (second order) are possible.
In the next section we explore these results in the context of
particular examples.

\section{Mean Field Solution - Particular Examples and Phase Diagrams}
\label{examples}

Consider the particular family of force functions
\begin{equation}
f(\phi)=-\phi (1+\phi^2)^m , \qquad g(\phi)=(1+\phi^2)^m,
\label{family}
\end{equation}
with $m\geq 1$.  We will concentrate in particular on the cases $m=1$
and $m=2$.

It is straightfoward to check that with this family of force functions,
for any value of $\Delta$ and coupling $D$, Eqs.~(\ref{ec3a}) have
only a single real solution, that is, each of the ordered systems is
pattern-free in the steady state.  Thus, any pattern observed in the
disordered system is purely a consequence of the disorder.

\begin{figure}
\begin{center}
\includegraphics[width = 8cm]{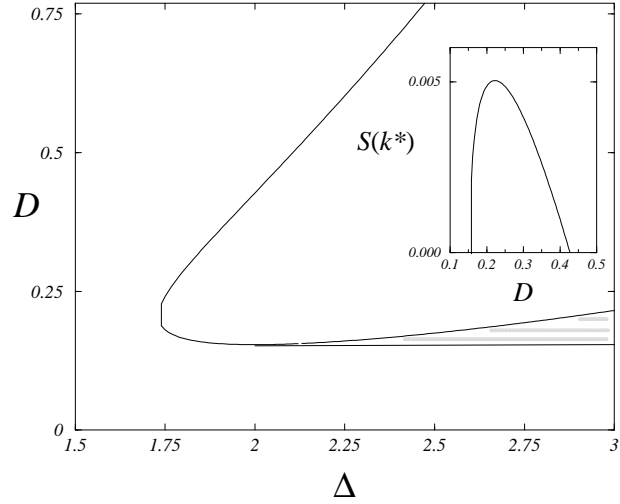}
\end{center}
\caption{Phase diagram for the case $m=1$. The shaded part of the
diagram is a coexistence region and the unshaded interior denotes the
occurrence of a single patterned phase. See text for a detailed
description.  The inset shows the behavior of the order parameter
defined in the text as a function of $D$ along the line $\Delta=2$.}
\label{f2a}
\end{figure}

We start with $m=1$, a choice that has been made in a number of studies
of purely noise induced transitions~\cite{chris,mangioni,kim}.
This is a particularly useful example because it can
be solved analytically in mean field. Furthermore, We solve the
coupled equations (\ref{thesearetheones}) fully and calculate the
resulting amplitude $\mathcal{A}$ using Eq.~(\ref{theproof}), check
the solutions for their stability properties, and thus construct a
phase diagram as a function of the parameters $\Delta$ and $D$.  
This procedure yields the diagram shown in Fig.~\ref{f2a}.
We will describe the features of the diagram in terms of the
analytic solutions to the problem as well as the general diagnostic
measures described in the previous sections.

Equations~(\ref{thesearetheones}) are cubic and yield altogether five
solutions.  One is the pattern-free solution $x_0=y_0$ for which
$\mathcal{A}(k^\ast)=0$.  Of the remaining four solutions, only two are
distinct (the other two are their negatives) and they yield:
\begin{equation}
\mathfrak{n}({k^\ast})\mathcal{A}_\pm(k^\ast) =\pm \frac{1}{2}\left(
\Delta^2 \pm \Delta\sqrt{2(D_1-2)} -\frac{3}{2}D_1 -1\right)^{1/2}.
\label{patterns}
\end{equation}
Note that in our equations we continue to use
$D_1$ for aesthetic reasons, whereas our illustrations involve $D$.
In two dimensions with the parameters specified earlier, the two are
related by $D_1=13D$.
Several features of these solutions are noteworthy and describe the
results in Fig.~\ref{f2a}:

\begin{figure}
\begin{center}
\includegraphics[width = 8cm]{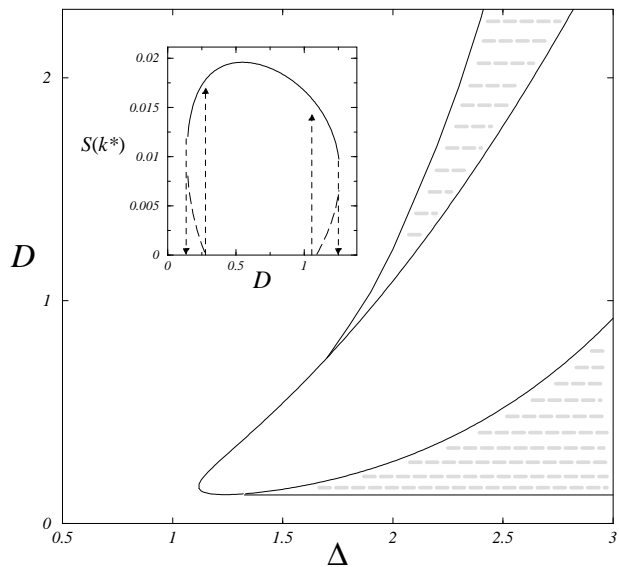}
\end{center}
\caption{Phase diagram for the case $m=2$.
As in the case $m=1$, the shaded portions represent coexistence regions and the unshaded interior indicates a single patterned phase. 
The inset graph shows  the order parameter as
a function of $D$ for $\Delta=2$. Note the
double-hysteresis behavior as a function of the coupling.}
\label{f3}
\end{figure}

1. Both solutions are complex if $D_1<2$. Therefore, the only stationary
state when $D_1<2$ is pattern-free (panels (a) in Figs.~\ref{f1}
and \ref{f2}).

2. When $D_1\geq 2$, both solutions are real in the parameter range
\begin{equation}
\Delta > \frac{1}{2}\left( \sqrt{2(D_1-2)} +2\sqrt{2D_1}\right).
\end{equation} 
One of these solutions ($\mathcal{A}_+$) is stable, the other unstable, and this
regime marks the shaded wedge in the figure, where the stable patterned
solution coexists with the pattern-free solution.  Entry into this
region marks a first order phase transition (panels (b) in
Figs.~\ref{f1} and \ref{f2}).  The two solutions merge (and the wedge
closes) at the point $\Delta=2$, $D_1=2$.

3. When $D_1\geq 2$ and
\begin{equation}
\frac{1}{2}\left(-\sqrt{2(D_1-2)} +2\sqrt{2D_1}\right)
<\Delta < \frac{1}{2}\left( \sqrt{2(D_1-2)} +2\sqrt{2D_1}\right),
\label{bounds}
\end{equation}
only the solution $\mathcal{A}_+$ is real.  This
solution is stable, and delimits
the unshaded region within which there is only a single steady state,
which is patterned.  The transition into this regime is part of the
first order phase transition if the crossing is from the shaded region,
or a second order transition if from the pattern-free region (panels (c)
in Figs.~\ref{f1} and \ref{f2}). The boundary of this regime is thus
precisely the curve defined by the bounds given in Eq.~(\ref{bounds}).
Alternatively, we can invert this equation and express these same bounds
as
\begin{equation}
D_1=\frac{2}{9} \left( 5\Delta^2 -3\pm 4\Delta \sqrt{\Delta^2-3}\right).
\label{exactly}
\end{equation}
Note that the two curves meet at the point $\Delta=\sqrt{3}$, $D_1=8/3$,
which is the leftmost point of the contour.

4. According to this description, the boundary of the unshaded region is
precisely the curve along which $F^\prime(x_0)=0$, thus bounding the
regime within which the pattern-free stationary state becomes unstable.
This is indeed the case, as we can establish without solving explicitly
for $x_0$ as a function of $D_1$ and $\Delta$
(the result is rather cumbersome). In terms of our original variable, for
our potential we have the explicit expression
$F^\prime(\phi)=-3\phi^2+2\Delta\phi +1$.  Setting this to zero yields
$\widetilde{\phi}_+= (\Delta \pm \sqrt{\Delta^2-1})/3$.  Requiring
that these values
indeed be ones that solve Eqs.~(\ref{thesearetheones}) with $x=y=x_0$
or, in our original notation, $F(\widetilde{\phi}_+)=D_1\widetilde{\phi}_+$,
again gives exactly Eq.~(\ref{exactly}).

We call special attention to the striking re-entrance behavior of
pattern formation as a function of the coupling $D$: sufficiently
strong coupling destroys any patterns.  Note that this implies that for
a given value of the disorder parameter $\Delta$ there is an optimal
coupling for which the patterned structure is most pronounced.  A more
nuanced discussion of this behavior requires quantification in terms of
order parameters.
In Appendix~\ref{b} we introduce in Eqs.~(\ref{sk})
and (\ref{j}) the \emph{total power spectrum} $S(k)$ and
the \emph{flux of convective heat} $J$. They are related to
one another in Eq.~(\ref{skj}), and they both contain useful information. 
In general, when there is no pattern at
all, $S(k)$ is independent of $k$ and of ${\mathcal O}(J/N)$, i.e.,
$S(k)/J={\mathcal O}(1/N)$ for all $k$.  On
the other hand, when there \emph{is} a pattern of wavevector magnitude
$k^\ast$, then $S(k^\ast)$ is of ${\mathcal O}(J)$, i.e., $S(k^\ast)/J
={\mathcal O}(1)$. A larger value of $S(k^\ast)$ indicates a stronger
pattern.  On the other hand, a larger ratio $S(k^\ast)/J$ indicates a
more coherent pattern.  It is possible (as we shall see below) for a
pattern to become more (or less) coherent even as it becomes less
(or more) pronounced.

In our mean field theory, however, we do not have access to all of this
information because we do not
deal with all modes but only with those that dominate
the pattern when one is present. We are therefore
restricted to choose as an order parameter
the value of the power spectrum at $k^\ast$:
\begin{equation}
S\left( k^{\ast }\right)
=\mathfrak{n}({k^\ast})\mathcal{A}^{2}\left( k^{\ast }\right).
\end{equation}
Within our theory $S\left( k^{\ast }\right)$  and $J$ are the same. 
When there is no pattern they are both zero and their ratio is
undefined.  When there is a pattern, $S(k^\ast)/J=1$.  Our theory can
therefore predict whether a pattern will become stronger as parameters
are modified, but not whether it will become more or less coherent.

The inset in Fig.~\ref{f2a} shows the power spectrum for the $m=1$ model
as a function of coupling at $\Delta =2$ obtained from our theory.
The values are just at
the edge of the second order phase transition region and, accordingly,
the order parameter rises continuously (albeit steeply) from zero upon
entry into the patterned region.  It also vanishes continuously as we
exit the patterned region, again indicating a second order transition
from the patterned state back to a pattern-free state (re-entrance) with
increasing coupling.  The most pronounced pattern is 
associated with the maximum in this curve, which can be found from
Eq.~(\ref{patterns}) to occur at $D_1=2.94$.

We have also carried out the same set of calculations for the potential
functions (\ref{family}) with $m=2$.  While some phase boundary
information can be obtained analytically, full analytic solution is no longer
possible because the equations are now quintic. A numerical solution
is straightforward and leads to the phase diagram shown in
Fig.~\ref{f3}.  The inset shows the order parameter for $\Delta=2$,
where the appearance and disappearance of patterns with increasing
coupling are \emph{both}
first-order transitions.  As a result, the order parameter
curve jumps discontinuously at \emph{both} ends and exhibits, at
both ends, the usual
hysteresis behavior associated with first-order transitions.

\section{Numerical Simulations}
\label{numerical}

In order to check the predictions of the modified mean field theory we
perform numerical simulations for the family of force functions
given in Eq.~(\ref{family})
with $m=1$ and $m=2$. We implement an Euler numerical scheme on a two
dimensional square lattice with periodic boundary conditions. The
values of the parameters used in the simulations are
$\Delta x=1$, $k_{0}=1$, $L=N\Delta x=64$, and $\Delta t=10^{-3}$.
The \emph{aspect ratio},
which measures the number of wavelengths of the expected patterns
that fit into the system, is
$L/\left( 2\pi /k^{\ast }\right) \sim 10$. Below we typically
present averages over ten disorder configurations.

\begin{figure}[htb]
\begin{center}
\epsfxsize = 3in
\epsffile{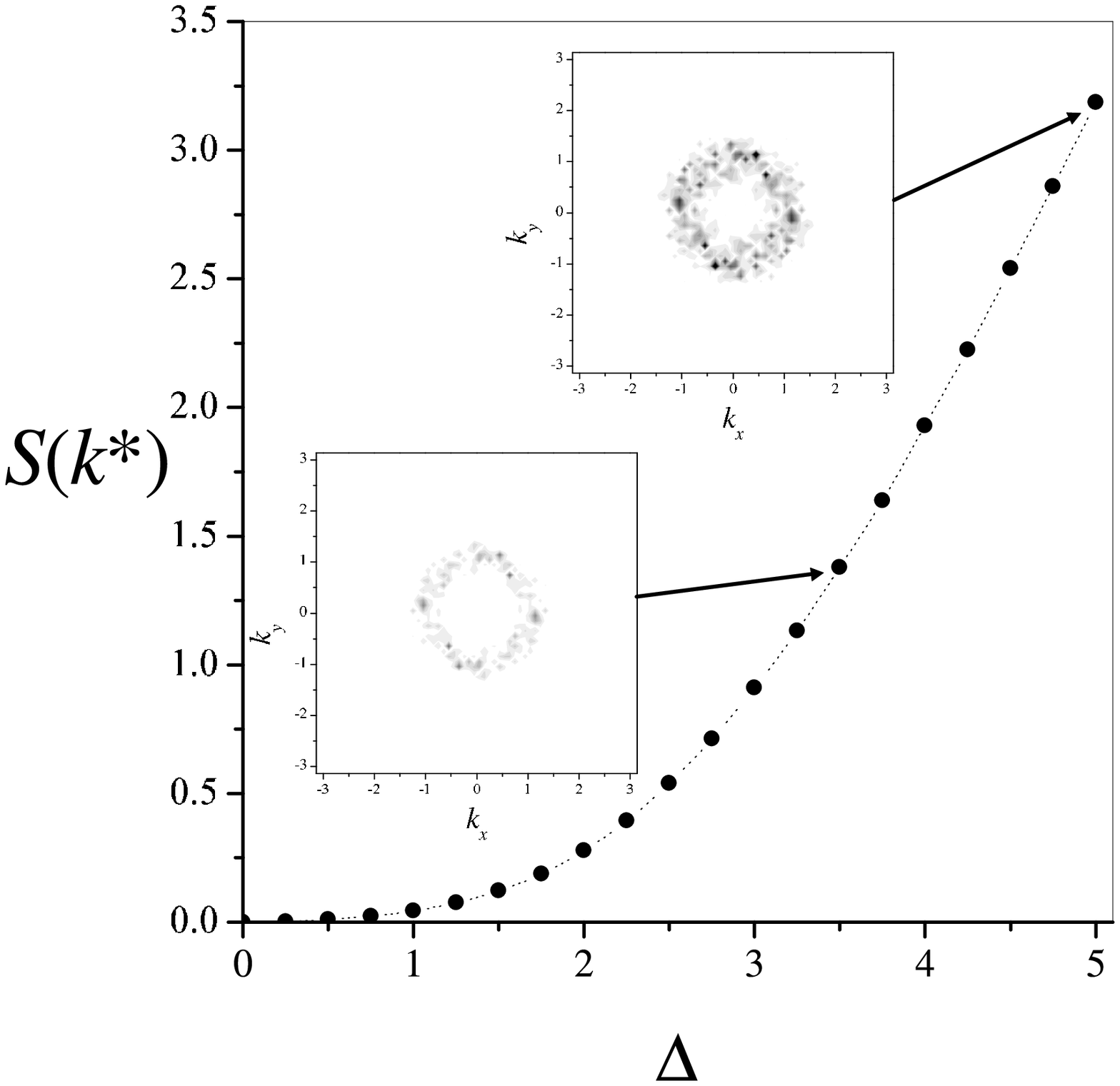}
\epsfxsize = 3in 
\epsffile{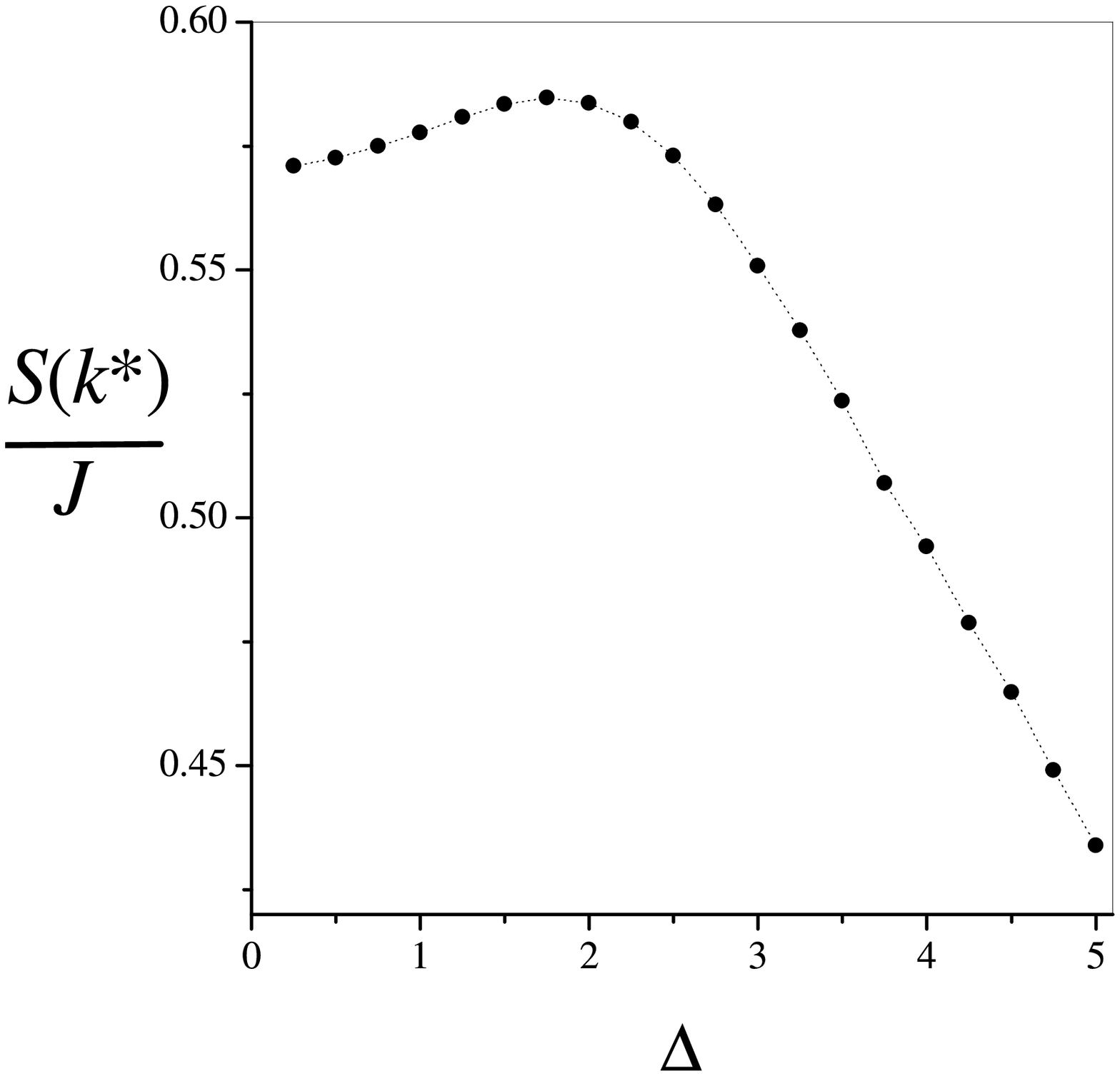}
\end{center}
\caption{First panel: $S(k^{\ast })$ vs $\Delta$ for $m=1$ and $D=5$. 
The insets are density plots of $\chi(\mathbf{k})$ for $\Delta=3.5$ and
for $\Delta=5$. Note
the destabilization of the $k^{\ast}$-modes as $\Delta$ grows.
The gray scale used in the density plot is the same for both insets.
Second panel:
$S(k^{\ast })/J$ vs $\Delta$. 
Note the nonmonotonic behavior, indicating maximal coherence at around
$\Delta=2$.}
\label{vsdeltaratio}
\end{figure}

Pattern formation is characterized by the total power spectrum, 
$S(k^\ast)$ [cf. Eq.~(\ref{sk})],
where the sum runs over wavevectors $\mathbf{k}$ whose magnitude
$k$ lies in the interval $[k^\ast-2\pi/L,k^\ast+2\pi/L]$.
In addition to this parameter and the flux of convective heat $J$,
another useful quantity for characterizing the system is the angular
average of $\chi \left( \mathbf{k}\right)\equiv
\widetilde{\phi}_{\mathbf{k}}\widetilde{\phi}_{-\mathbf{k}}$.

The first panel in
Fig.~\ref{vsdeltaratio} shows the order parameter $S(k^{\ast })$
as a function of the disorder intensity parameter $\Delta$ for the
family $m=1$ and for coupling coefficient $D=5$.  In the absence of
disorder there is no pattern, $S(k^{\ast })\approx 0$, but a pattern clearly
develops as $\Delta$ increases and is therefore entirely due to the
disorder.  The inset graphs show
$\chi\left( \mathbf{k}\right)$ by means of a wavevector density plot
for $\Delta =3.5$ and for $\Delta =5$. Clearly, a ring of unstable
modes develops around $k^{\ast }$, and the ring becomes more prominent
with increasing disorder (as measured by the value of $\Delta$).  While we
recognize that our modified mean field theory does not predict the
transition values quantitatively (for $D=5$ we predict a patterned state
to first appear when $\Delta=11.5$ while the simulations already show
pattern formation for much smaller values of $\Delta$), the qualitative
behavior is as predicted. 
It should be noted that while the intensity of
$\chi(\mathbf{k})$ at the most unstable modes increases as indicated by
the gray scale, the {\em width} of
the ring around $k^\ast$ also increases with increasing $\Delta$.
While a higher intensity indicates a more pronounced pattern, the width is
a measure of the {\em coherence} of the patterns, increasing width
indicating greater decoherence.  As mentioned earlier,
the ratio $S(k^\ast)/J$ can be used to characterize
the coherence of the patterns. 
This nonmonotonic ratio is shown in the second panel in
Fig.~\ref{vsdeltaratio}. As also noted earlier,
this ratio can not be obtained from our mean field theory.

\begin{figure}[htb]
\begin{center}
\epsfxsize = 3in
\epsffile{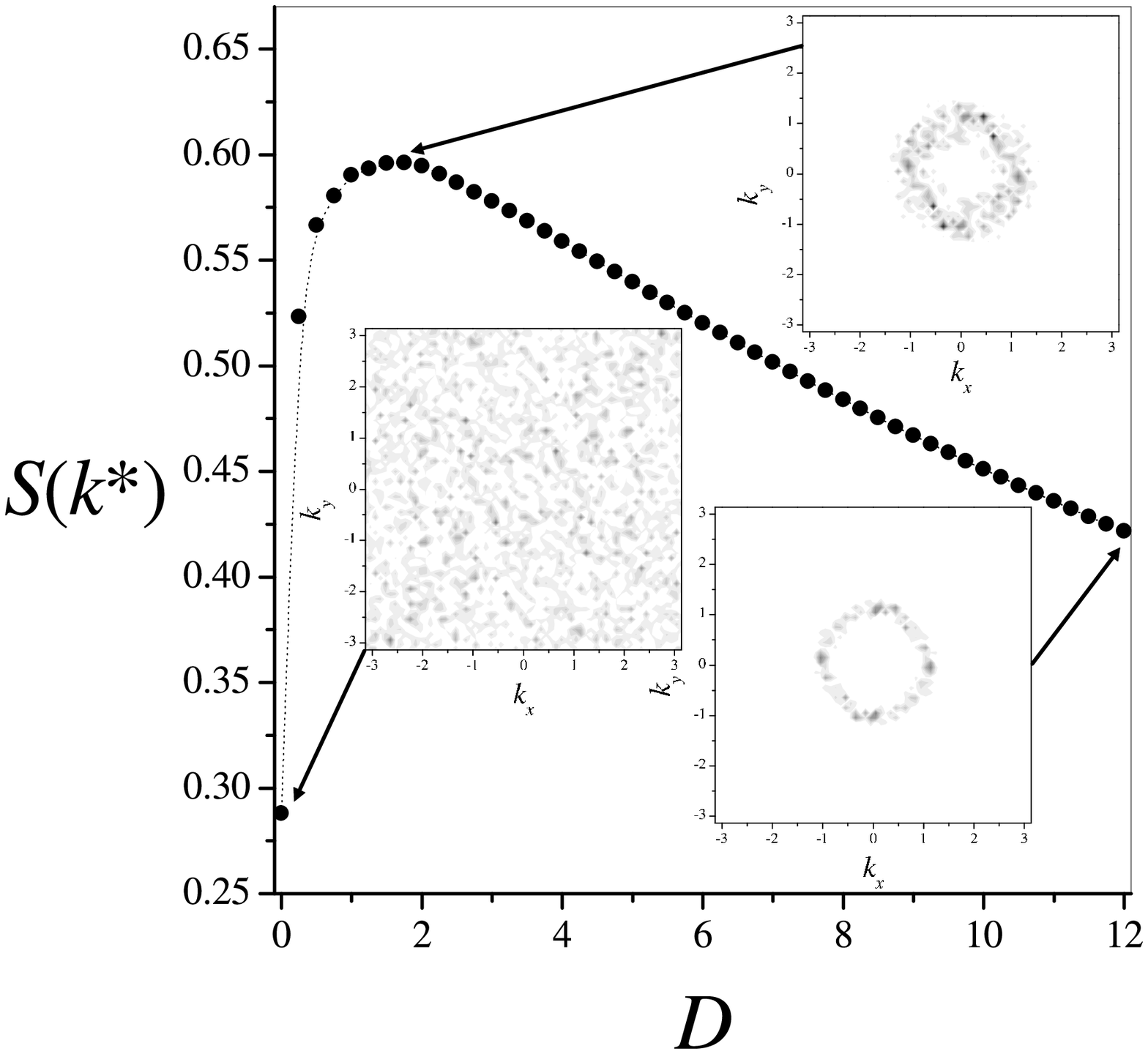}
\epsfxsize = 3in 
\epsffile{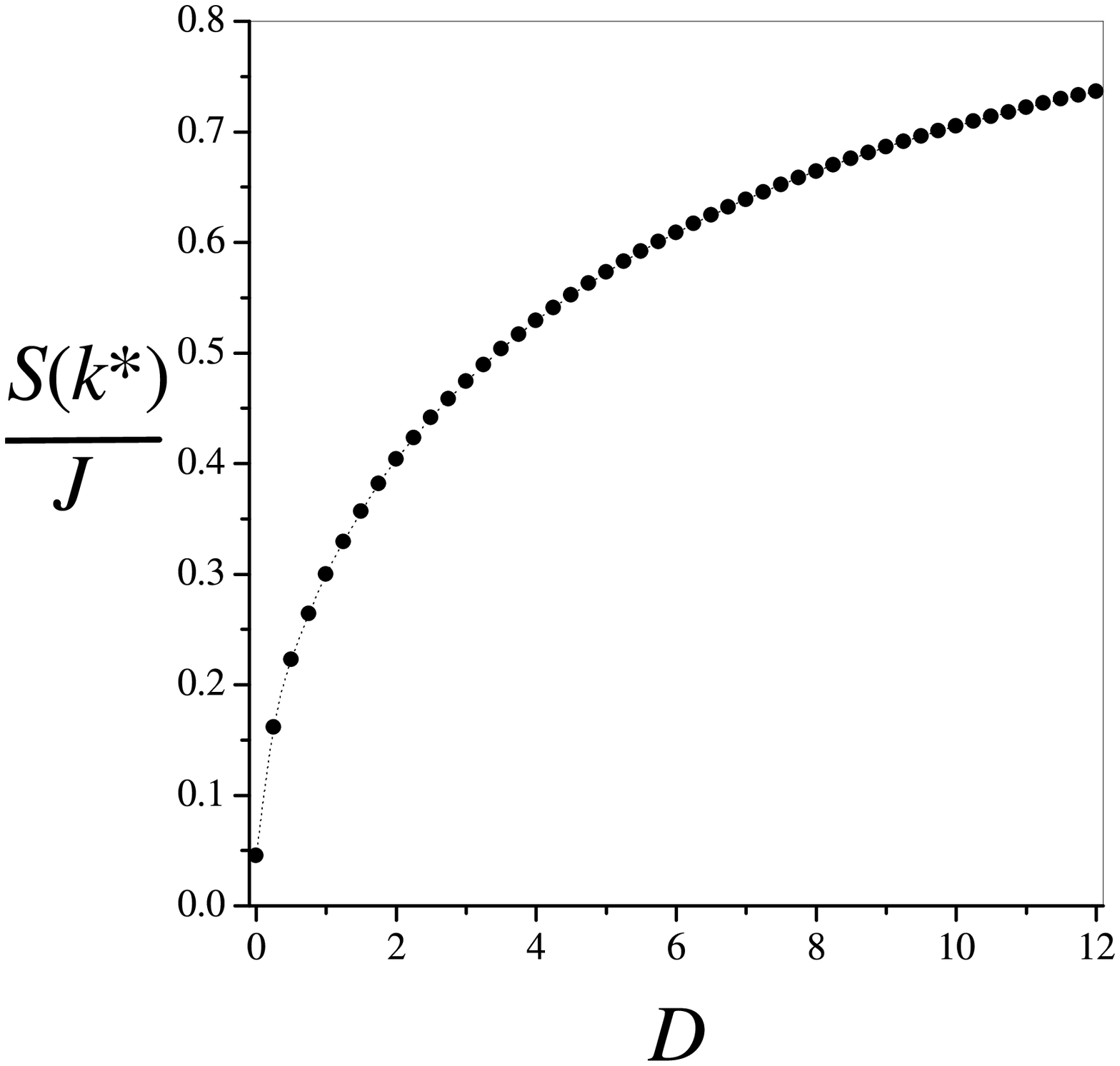}
\end{center}
\caption{First panel: re-entrant behavior of the total power spectrum as a
function of the coupling for $m=1$ and $\Delta=2.5$. The inset panels
show $\chi(\mathbf{k})$ for $D=0$, $D=2$,
and $D=12$ by means of density plots: the same gray scale is used
in all cases.  Although $S(k^\ast)$ shows non-monotonic behavior,
the pattern coherence increases as the coupling grows, as seen in the
monotonic behavior of the ratio $S(k^{\ast })/J$ in the second panel.}
\label{vsdif1ratio2}
\end{figure}

The re-entrant behavior as a function of the coupling
predicted by the modified mean field theory
is shown in the first panel of
Fig.~\ref{vsdif1ratio2} for $m=1$ and $\Delta =2.5$.
The inset panels depict density plots of $\chi(\mathbf{k})$
for $D=0$, $D=2$, and $D=12$. Again, the agreement with the theory is
only qualitative but nevertheless dramatic because re-entrance with increasing
coupling strength is a rare phenomenon. Notice the destabilization of a ring
of modes around $k^{\ast }$ with increasing coupling, and its
subsequent intensity fade-out as $D$ increases further.
It should be noted that not only does the intensity of
$\chi(\mathbf{k})$ at the most unstable modes decrease as indicated by
the gray scale, but the width of
the ring also shrinks around $k^{\ast }$ as the coupling grows.
The patterns thus become less pronounced but more coherent with increasing
coupling.
A representation of $S\left( k^{\ast }\right) /J$, shown in
the second panel of Fig.~\ref{vsdif1ratio2}, reveals
that although the system presents a re-entrant behavior with the coupling,
the coherence actually increases monotonically as a function of $D$.

\begin{figure}
\includegraphics[width = 5cm]{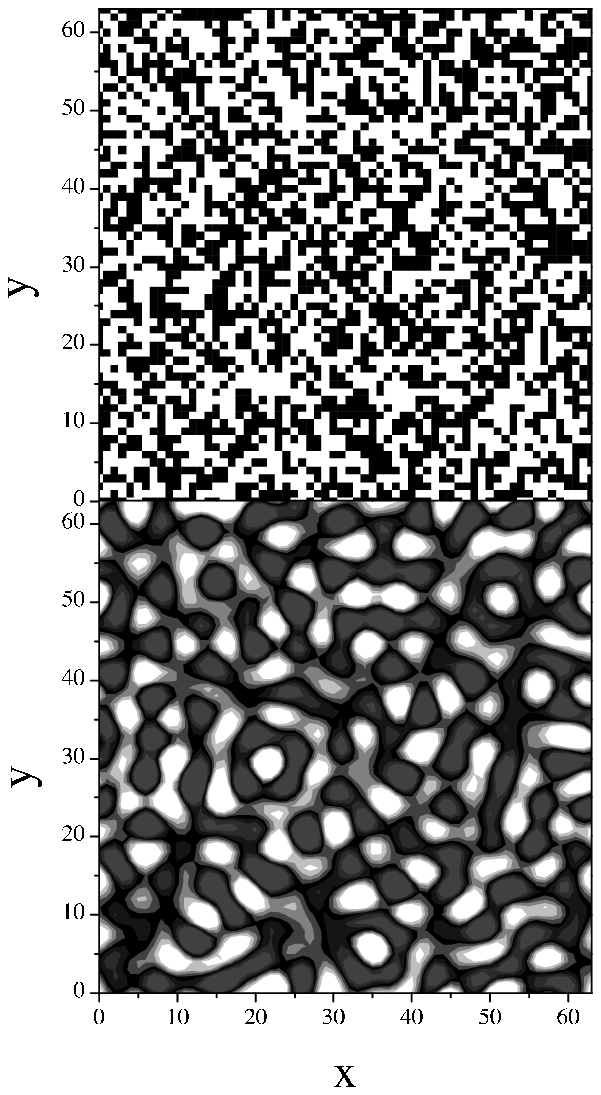}
\caption{Density plot showing a stationary pattern purely induced by
disorder for $m=1$, $\Delta=2.5$, and $D=12$ (lower panel). The
upper panel shows the underlying configuration of quenched disorder for
this particular simulation.}
\label{pattern}
\end{figure}

The actual stationary spatial patterned configuration induced by the 
disorder is shown in the lower panel of Fig.~\ref{pattern}
for $m=1$, $\Delta =2.5$ and $D=5$,
i.e., the rightmost point in the first panel of
Fig.~\ref{vsdif1ratio2}. The upper panel
in Fig.~\ref{pattern} shows the particular configuration of
quenched disorder in the simulation that leads to the stationary field
shown in the lower panel.

\begin{figure}
\includegraphics[width = 8cm]{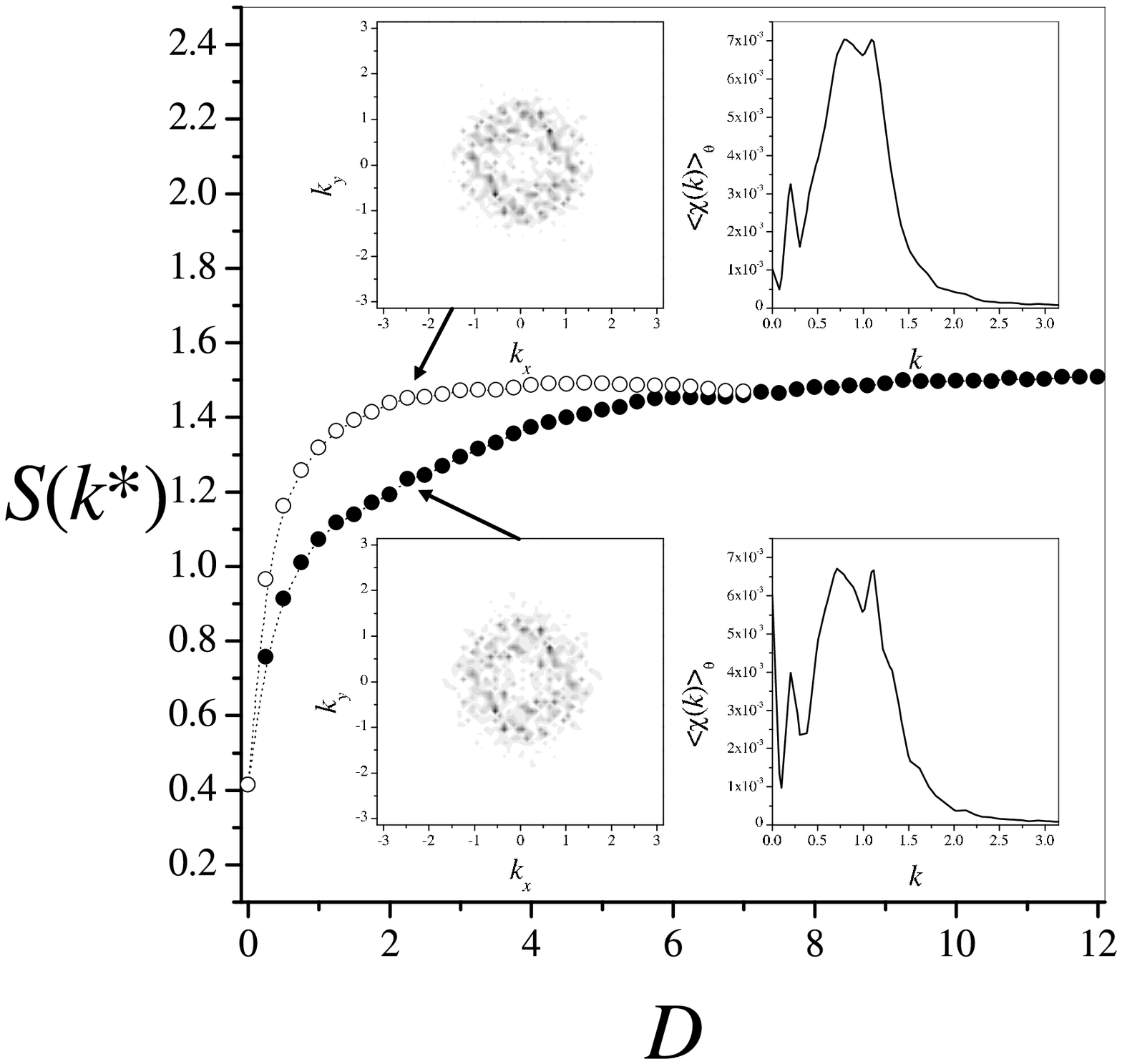}
\caption{Total power spectrum as a function of the coupling for $m=2$
and $\Delta=3$. The hysteresis is reflected in the difference between
the black and white circle phase space trajectories, obtained when
increasing and decreasing $D$ respectively. The insets show density plots of
$\chi \left(\mathbf{k}\right)$ and their angular averages.}
\label{vsdif2}
\end{figure}

We have thus confirmed the disordered-induced pattern formation
phenomenon but have not yet ascertained our other prediction, namely,
the occurrence of hysteresis in some cases.  We predict hysteresis to
occur in the shaded regions of the phase diagrams in Figs.~\ref{f2a}
and~\ref{f3}.  Although we do not necessarily expect
quantitative agreement with the particular values of parameters that
lead to hysteresis, the question is whether hysteresis is observed at
all.

Hysteresis implies a memory of the initial conditions. To detect
hysteresis we carry out two simulations.  In one, starting from an
unpatterned initial condition $\phi _{\mathbf{r}}(t_0)=0$ for all
$\mathbf{r}$,
we calculate the total power spectrum in the steady state as a
function of increasing coupling.  Then we decrease the coupling along
the same phase space path, but now our initial condition for
each value of the coupling is the stationary state obtained in the
simulation with the previous value of $D$.  A difference in the
spectrum obtained for these two different
conditions reflects hysteresis and the attendant coexistence of different
stationary states (one patterned and one unpatterned, as discussed
earlier).  In Fig.~\ref{vsdif2} we observe precisely this behavior
for the family of functions $m=2$
when increasing (black circles) and decreasing (white circles) the coupling
for a fixed value of the intensity of the disorder,
$\Delta =3$. Hysteresis is observed between $D=0$ and $D\sim 7$. The
qualitative behavior is thus exactly as predicted by the mean field
model. 

The coexistence of patterned and unpatterned states is discernible in
the density plot insets of $\chi\left(\mathbf{k}\right)$
in the figure in that they are no longer cleanly
annular but now include contributions from wavevectors other than those
of magnitude near $k^\ast$.  The difference between the two insets
(indicating different relative contributions of patterned and
unpatterned states) is visible but, an even clearer rendition of the
difference is seen in the angular average 
$\left\langle \chi(\mathbf{k})\right\rangle_{\theta }$ shown in the
other two inset panels.  Note, for instance, that the contribution of
modes near $k=0$ are relevant when increasing the coupling
but very small when decreasing $D$.

\section{Summary}
\label{conclusions}
Using a modified mean field theory, we have explored 
general conditions under which
spatially extended systems with coupling {\em a la} Swift-Hohenberg
exhibit pattern
formation {\em purely} induced by the presence of quenched dichotomous
disorder.
We have illustrated the phenomenology with a family of force functions that
includes the paradigmatic models of noise-induced phase transitions
and noise-induced spatial patterns, among them one that can fully
be solved analytically within the mean field model.
We show that pattern formation can be achieved through continuous
(second order) and discontinuous (first order)
transitions, and that the pattern formation phenomenon is re-entrant
with the coupling.  Thus, increasing the coupling eventually destroys
the order.  All these predictions have been checked by means of
numerical simulations, and we find full qualitative (albeit not
quantitative) agreement with the theory.

Beyond the capabilities of the modified mean field approach, we have
explored not only the occurrence of patterns but also their coherence.
The numerical simulations show that even while patterns become stronger
(as reflected by a larger value of the total power spectrum at the
particular wavevector magnitude characteristic of the pattern), they may
become less coherent (as reflected by a larger contribution of
neighboring wavevectors).  The converse is also possible: patterns that
become more coherent even as they weaken.

We end by stressing that the methodology and formalism developed herein
can be generalized straightforwardly to other kinds of
disorder and/or couplings.

\section*{Acknowledgments}
We thank J. M. R. Parrondo for fruitful discussions.
This work was supported by the Engineering Research Program of
the Office of Basic Energy Sciences at the U. S. Department of Energy
under Grant No. DE-FG03-86ER13606. Support was also
provided by MECD-Spain Grant No. EX2001-02880680, and by MCYT-Spain Grant
No. BFM2001-0291.

\appendix

\section{Generalization of Modified Mean Field Theory}
\label{a}

We begin by illustrating some detailed dependences associated with the
ansatz field~(\ref{mean field}) and the action of ${\mathcal L}$ on it.
For example, for an ${\bm r}'$ that
is $m$ lattice sites away from ${\bm r}=(x_1, x_2, \ldots, x_d)$
in the direction $j$ the ansatz reads
\begin{equation}
\phi(x_1,x_2,\ldots,x_j+m\Delta x,\ldots,x_d)
=\sum_{\left\{\bm{k}^{\ast}\right\}}\mathcal{A}(k^*)
\cos(m\Delta x k_j) +B.
\label{axis}
\end{equation}
For an  ${\bm r}'$ that is in the immediate positive diagonal location
away from ${\bm r}$ we have
\begin{equation}
\phi(x_1+\Delta x,x_2+\Delta x,\ldots,x_j+\Delta x,\ldots,x_d +\Delta x)
=\sum_{\left\{\bm{k}^{\ast}\right\}}\mathcal{A}(k^*)
\cos[\Delta x (k_1 +k_2+\cdots+k_d)] +B.
\label{diagonal}
\end{equation}
Next, to apply the discrete version (\ref{l-discrete}) of ${\mathcal L}$
we must elucidate
the effect of the operators
$\left[ \sum_{i=1}^{d}\sinh^2\left(\frac{\Delta x}{2}\frac{\partial}
{\partial x_i}\right)\right]^n$ on the field
$\phi_{\bm{r}}$ for $n=1,2$. With $n=1$, we use the relation
$2\sinh^2(y/2)=
[\cosh(y)-1]$ and note that
\begin{equation}
\begin{aligned}
\sum_{i=1}^d\cosh\left(\Delta x\frac{\partial}{\partial x_i}\right)
\phi_{\bm{r}}  =&\frac{1}{2}\left[\phi(x_1+\Delta
x,x_2,\ldots,x_j,\ldots,x_d) \right. \\
&~~~~ +\phi(x_1-\Delta x,x_2,\ldots,x_j,\ldots,x_d) \\ \\
& \ldots+\phi(x_1,x_2,\ldots,x_j+\Delta x,\ldots,x_d) \\ \\
&~~~~+\phi(x_1,x_2,\ldots,x_j-\Delta x,\ldots,x_d)  \\ \\
& \ldots+\phi(x_1,x_2,\ldots,x_j,\ldots,x_d+\Delta x)\\ \\
&\left. ~~~~+\phi(x_1,x_2,\ldots,x_j,\ldots,x_d-\Delta x)
\right] .
\end{aligned}
\end{equation}
By using Eq.~(\ref{axis}) in this last equation we obtain
\begin{equation}
\sum_{i=1}^d\cosh\left(\Delta x\frac{\partial}{\partial x_i}\right)
\phi_{\bm{r}}=\sum_{\left\{ \bm{k}^{\ast}\right\}}\mathcal{A}(
k^*)\sum_{i=1}^d\cos(k_i\Delta x) + B.
\label{cosh}
\end{equation}
As for $n=2$, we note that
$4\sinh^2(y/2)\sinh^2(z/2)=[\cosh(y)-1][\cosh(z)-1]$ and, in turn,
$\cosh(y)\cosh(z)=\frac{1}{2}[\cosh(y+z)+\cosh(y-z)]$.  The latter
combination leads to contributions that involve
both forward and backward
translations in different spatial directions. This is easily
visualized by noting explicitly that
\begin{equation}
\begin{aligned}
\left[\sum_{i=1}^d\cosh\left(\Delta x\frac{\partial}{\partial x_i}
\right) \right] ^{2}  =&\frac{1}{2}\left[\sum_{i,j=1}^{d}\cosh\left(
\Delta x\left(\frac{\partial}{\partial x_i}+\frac{\partial}{\partial
x_j}
\right) \right) \right. \\
& + \left. \cosh\left( \Delta x\left(\frac{\partial}{\partial
x_i}-\frac{\partial}{\partial x_j}\right) \right) \right] .
\end{aligned}
\end{equation}
Notice that for the $d$ cases where with $i=j$, the second term on the
right
hand side leaves the field at the original site $\bm{r}$. The field at
the
original site is not represented by the ansatz assumption, and therefore
we must subtract the
$d$ ``spurious" terms produced by the ansatz state
and add $d$ times the field $\phi_{\bm{r}}$. This procedure leads to,
\begin{equation}
\left[ \sum_{i=1}^d\cosh\left(\Delta x\frac{\partial}{\partial x_i}
\right) \right]^{2}\phi_{\bm{r}}=\frac{d}{2}\phi_{\bm{r}}
+\sum_{\left\{\bm{k}^{\ast}\right\} }\mathcal{A}(k^*)
\left[ \left(\sum_{i=1}^d\cos(k_i\Delta x)
\right)^{2}-\frac{d}{2}\right] -\frac{d}{2} B .
\label{cosh2}
\end{equation}
Note that we have taken advantage of the directional insensitivity of
$k^*$.

Use of Eqs.~(\ref{cosh}) and (\ref{cosh2}) in
Eq.~(\ref{l-discrete})
then leads to the following approximation for the term containing the
Swift-Hohenberg coupling operator:
\begin{equation}
\mathcal{L}\phi_{\bm{r}}=D_1\left( \sum_{\left\{\bm{k}^{\ast
}\right\} }\mathcal{A}(k^*) -\phi_{\bm{r}}\right)
+ B(D_1-Dk_0^4),
\label{couplingapprox}
\end{equation}
where $D_1$ is given in Eq.~(\ref{d1}).

Finally, since the summand in Eq.~(\ref{couplingapprox}) is
independent of the direction
of the ${\bm k}^*$, the sums simply give the number of terms
$\mathfrak{n}(k^{\ast})$ in the sum, as given in Eq.~(\ref{frakn})
(or the appropriate integral form), times the summand.  
Thus we finally arrive at the mean field approximation
\begin{equation}
\mathcal{L}\phi_{\bm{r}}=D_1\left[\mathfrak{n}(k^{\ast})
\mathcal{A}(k^{\ast})-\phi_{\bm{r}}\right] +B(D_1-Dk_0^4).
\end{equation}

\section{Normalization of the Fourier Transform and Order Parameter}
\label{b}

In this appendix we provide details on the relations between different
relevant parameters used in the characterization of pattern formation.

The Fourier transform of a field $\phi _{\mathbf{r}}$ and its inverse read
respectively 
\begin{eqnarray}
\widetilde{\phi}_{\mathbf{k}} &=&C\sum_{\mathbf{r}}
\phi _{\mathbf{r}}e^{-i\mathbf{k\cdot r}},  \label{ft} \\
\phi _{\mathbf{r}} &=&\widetilde{C}\sum_{\mathbf{k}}
\widetilde{\phi}_{\mathbf{k}}e^{i\mathbf{k\cdot r}},  \label{fti}
\end{eqnarray}
where $C$ and $\widetilde{C}$ are normalization constants. Since
\begin{eqnarray}
\frac{1}{N^{d}}\sum_{\mathbf{k}}e^{i\mathbf{k} \cdot
\left( \mathbf{r}-\mathbf{r}^{\prime }\right) }
&=&\delta _{\mathbf{r},\mathbf{r}^{\prime }}, \\
\frac{1}{N^{d}}\sum_{\mathbf{r}}e^{i\mathbf{r} \cdot
\left( \mathbf{k}-\mathbf{k}^{\prime }\right) }
&=&\delta _{\mathbf{k},\mathbf{k}^{\prime }},
\end{eqnarray}
the functional relation between $C$ and $\widetilde{C}$ can be obtained
by substituting Eq. (\ref{ft}) into Eq. (\ref{fti}).  One readily obtains 
\begin{equation}
\widetilde{C}=\frac{1}{CN^d} .
\end{equation}
For simplicity we choose $C=1/N^{d}$, and therefore $\widetilde{C}=1$.

Two parameters commonly used to characterize spatial patterns are the
\emph{total power spectrum}, $S\left( k\right) $,
and the \emph{flux of convective heat}, $J$, 
\begin{eqnarray}
S\left( k\right) &=&\sum_{\left\{k\right\} }
\widetilde{\phi}_{\mathbf{k}}\tilde{\phi}_{-\mathbf{k}},
 \label{sk} \\
J &=&\frac{1}{N^{d}}\sum_{\mathbf{r}}\phi _{\mathbf{r}}^{2},
\label{j}
\end{eqnarray}
where the sum in (\ref{sk}) runs over all modes of magnitude $k$. It is easy
to check that the functional relation between these two quantities is simply 
\begin{equation}
J=\sum_{k}S\left( k\right) ,
\label{skj}
\end{equation}
where now the sum runs over the magnitude of the modes.

For a \emph{pure} spatial pattern of wave vector
of magnitude $k^{\ast}$
where all spatial directions contribute in the same way we expect
$\phi _{\mathbf{r}}$ to be 
\begin{equation}
\phi _{\mathbf{r}}=\sum_{\left\{k^{\ast }\right\}} 
\mathcal{A}\left( k\right) \cos \left( \mathbf{k\cdot r}\right).
\label{field}
\end{equation}
Therefore, the Fourier transform of such a field is 
\begin{equation}
\widetilde{\phi}_{\mathbf{k}}=\frac{1}{2}\mathcal{A}\left( k^{\ast }\right)
\sum_{\left\{k^{\ast}\right\}}
\left( \delta _{\mathbf{k},\mathbf{k}^{\ast }}
+\delta _{-\mathbf{k},\mathbf{k}^{\ast }}\right) .  
\end{equation}
Consequently, the total power spectrum and the flux of this field are
respectively, 
\begin{eqnarray}
S\left( k\right)  &=&\frac{1}{4}\sum_{\left\{ k\right\} }
\mathcal{A}^{2}\left( k^{\ast }\right)
\sum_{\left\{k^{\ast}\right\}}
\left( \delta _{\mathbf{k},\mathbf{k}^{\ast
}}+\delta _{-\mathbf{k},\mathbf{k}^{\ast }}\right) ^{2}=\mathfrak{n}\left(
k^{\ast }\right) \mathcal{A}^{2}\left( k^{\ast }\right)
\delta _{k,k^{\ast }}. \\
J &=&\mathfrak{n}\left( k^{\ast }\right) \mathcal{A}^{2}\left( k^{\ast }\right).
\end{eqnarray}

\end{document}